\documentclass[%
reprint,
superscriptaddress,
 amsmath,amssymb,
 aip, pop,
floatfix,
]{revtex4-2}

\usepackage{graphicx}
\usepackage{dcolumn}
\usepackage{bm}
\usepackage{hyperref}

\begin{document}

\title{Numerical study of Langmuir wave coalescence in laser-plasma interaction}

\newcommand{\LULI}{\affiliation{LULI, CNRS, CEA, Ecole Polytechnique, Institut Polytechnique de Paris, 91128 Palaiseau cedex, France}}
\newcommand{\LESIA}{\affiliation{LESIA, Observatoire de Paris-PSL, CNRS, Sorbonne Universit\'e, Universit\'e de Paris, 92195 Meudon, France}}
\newcommand{\DIF}{\affiliation{CEA, DAM, DIF, F-91297 Arpajon, France}}
\newcommand{\LMCE}{\affiliation{Universit\'e Paris-Saclay, CEA, LMCE, 91680 Bruy\`eres-le-Ch\^atel, France}}

\author{F.~P\'erez}
\email{frederic.perez@polytechnique.edu}
\LULI

\author{F.~Amiranoff} \LULI
\author{C.~Briand} \LESIA
\author{S.~Depierreux} \DIF
\author{M.~Grech} \LULI
\author{L.~Lancia} \LULI
\author{P.~Loiseau} \DIF \LMCE
\author{J.-R.~Marqu\`es} \LULI
\author{C.~Riconda} \LULI
\author{T.~Vinci} \LULI

\date{\today}

\begin{abstract}
Type-III-burst radio signals can be mimicked in the laboratory via laser-plasma interaction. Instead of an electron beam generating Langmuir waves (LW) in the interplanetary medium, the LWs are created by a laser interacting with a millimeter-sized plasma through the stimulated Raman instability. In both cases, the LWs feed the Langmuir decay instability which scatters them in several directions. The resulting LWs may couple to form electromagnetic emission at twice the plasma frequency, which has been detected in the interplanetary medium, and recently in a laboratory laser experiment [Marquès \textit{et al.} Phys. Rev. Lett. 124, 135001 (2020)]. This article presents the first numerical analysis of this laser configuration using particle-in-cell simulations, providing details on the wave spectra that are too difficult to measure in experiments. The role of some parameters is addressed, with a focus on laser intensity, in order to illustrate the behavior of the electromagnetic emission's angular distribution and polarization.
\\

The following article has been accepted by Physics of Plasmas.
\\

Copyright (2021) F. P\'erez, F. Amiranoff, C. Briand, S. Depierreux, M. Grech, L. Lancia, P. Loiseau, J.-R. Marqu\`es, C. Riconda, T. Vinci. This article is distributed under a Creative Commons Attribution (CC BY) License.
\end{abstract}

\maketitle

\newcommand{\LW}{\textrm{LW}}
\newcommand{\LWC}{{2\omega_p}}
\newcommand{\IAW}{\textrm{IAW}}

\section{Introduction\label{sec-intro}}

Electron beams from solar eruptions or interplanetary shocks produce intense radio emissions \cite{pick2008,reid2014} at the plasma frequency $\omega_p$ and its harmonic $2\omega_p$. Following the original idea of Ginzburg \& Zhelezniakov \cite{ginzburg1958}, the beams provide the free energy for the bump-on-tail instability which generates Langmuir waves (LW). The decay of these plasma waves leads to the emission of the electromagnetic waves. In particular, we study a two-step process for the emission at $2\omega_p$: it relies on the Langmuir decay instability (LDI) \cite{zakharov1972,robinson1997} which scatters LWs on ion acoustic waves (IAW); the resulting, counter-propagating LWs couple with the forward LWs to form an electromagnetic (EM) wave at $\sim2\omega_p$.

In the interplanetary medium, experimental spacecraft measurements have evidenced this coupling \cite{henri2009} and provided characterization of the EM source location \cite{baumback1976,bougeret1984,cecconi2005}. However, the EM emissions are subjected to scattering effects in the plasma cloud, broadening the apparent source size \cite{krupar2014,kontar2019}. The question is thus to estimate the relative contribution of the scattering and the intrinsic emission pattern in the source size, as observed by spacecraft. Laboratory measurements may add complementary insights, and a first laser-plasma interaction experiment has recently provided data on $2\omega_p$ emission \cite{marques2020}. Instead of an electron beam, a high-energy nanosecond laser produces LWs by stimulated Raman scattering (SRS). They can initiate the LDI \cite{depierreux2000a} and generate EM emission at $2\omega_p$. The experimental results supported the above scenario for the $2\omega_p$ emission, and that its polarization is aligned with the plane containing the laser axis and the direction at which the emission is observed.

The present article explores this laser setup via large-scale numerical particle-in-cell (PIC) simulations. They allow to perform parametric studies and provide direct information on electrostatic waves that was not accessible in the experiment. These simulations support that the LDI-based scenario is dominant in these conditions, with conversion efficiencies and polarization coherent with the experimental results of Ref. \onlinecite{marques2020}. By varying several parameters, we show that the $2\omega_p$ emission's angular distribution and polarization are strongly influenced by the growth of the LDI. Indeed, the LDI tends to scatter LWs at large angles which imprint their direction on the $2\omega_p$ EM waves, thus reducing the $2\omega_p$ degree of polarization. The important parameters are mostly the wave amplitudes and the plasma density (other parameters do matter, e.g. temperature and mass ratios, but they are constrained by the experimental conditions). Thus we focus on a comparison between low-power and high-power laser interaction to exemplify the different outcomes in terms of $2\omega_p$ emission. Note that previous numerical works have explored the growth of LDI in laser-plasma interaction using reduced models for selected processes\cite{russel1999}. Even though the plasma conditions were somewhat different, we obtain similar trends in the present article, while our numerical modeling relies on a first-principles approach. In addition, we address the production of $2\omega_p$, which was not previously considered.

Section \ref{sec-processes} recalls the wave processes that result in $2\omega_p$ emission. In section \ref{sec-setup}, we summarize the relevant experimental configuration and detail our simulation setup. Sections \ref{sec-loP} and \ref{sec-hiP} are devoted to the cases of a low-power and a high-power laser, respectively. We discuss these findings with respect to the experimental results in section \ref{sec-comparison} and consider the relevance of laser experiments for space scenarios in section \ref{sec-conclusion}.

\section{Basic processes\label{sec-processes}}

All notations are given in Appendix~\ref{app-notations}. We recall the dispersion relations: $\omega^2=\omega_p^2+k^2c^2$ for electromagnetic waves, $\omega^2=\omega_p^2+3k^2 v_\textrm{th}^2$ for LWs and $\omega^2\sim k^2 c_s^2$ for IAWs.

Here, the generation of radiation at $2\omega_p$ can be described in three steps. First, the SRS instability scatters and redshifts an EM wave from the laser, off one LW:
\begin{equation*}
    \textrm{EM}_\textrm{laser} \rightarrow \textrm{EM}_\textrm{SRS} + \LW_0
\end{equation*}
As the LW frequency is close to $\omega_p$ the scattered EM wave frequency is close to $\omega_0-\omega_p$ to ensure energy conservation. The conservation of momentum then gives the following wave vector for the LW:
\begin{equation}
    \label{eq-ramanLW}
    k_{\LW0} = \left( \cos\alpha +\sqrt{\cos^2\alpha-\eta\frac{2-\eta}{1-\eta^2}} \right) k_0
\end{equation}
where $k_0$ is the laser wave-vector in the plasma, $\eta=\omega_p/\omega_0$ and $\alpha$ is the angle between $\vec k_{\LW0}$ and $\vec k_0$. Note that the fastest-growing mode corresponds to backward-scattering ($\alpha=0$), with the scattered EM wave going backward and the LW going forward\cite{kruer}.

The second step, in the regime of low $k_{\LW}\lambda_D$ (this parameter determining Landau damping), consists in the LDI where the previous LW is scattered by an IAW:
\begin{equation*}
    \LW_0 \rightarrow \LW_1 + \textrm{IAW}_1
\end{equation*}
The IAW carries very little energy so that the secondary LW has almost the same frequency, and a slightly decreased wave-vector\cite{depierreux2000b}. The conservation of momentum gives the following wave-vectors for the secondary LW and for the IAW:
\begin{eqnarray}
    \label{eq-LDILW}
    k_{\LW1} \sim k_{\LW0} - \frac{2 c_s \omega_p}{3 v_\textrm{th}^2} \sin\frac\theta 2\\
    \label{eq-LDIIAW}
    k_{\IAW1} \sim 2k_{\LW0}\cos\varphi - \frac{2 c_s \omega_p}{3 v_\textrm{th}^2}
\end{eqnarray}
where $\theta$ is the angle between $\vec k_{\LW1}$ and $\vec k_{\LW0}$, and $\varphi$ is the angle between $\vec k_{\IAW1}$ and $\vec k_{\LW0}$. The fastest growing mode corresponds to the LW being scattered backwards, but with a large angular acceptance (it requires multi-dimensional treatment). Note that this process can be repeated on LW$_1$ to produce yet another LW, and so on, creating an LDI cascade\cite{briand2014,robinson1997}.

The third step (LW coalescence) results from the nonlinear current produced by the beating of two LWs which yields an EM wave at $2\omega_p$:
\begin{equation*}
    \LW_i + \LW_j \rightarrow \textrm{EM}_\LWC
\end{equation*}
The dispersion relation gives $k_{\LWC}\sim\sqrt{3}\,\omega_p/c$.
The polarization of the EM wave is in the plane given by the wavevectors $\vec k_{\LW i}$ and $\vec k_{\LW j}$.

\section{Experimental and numerical setup\label{sec-setup}}

The experiment from Ref. \onlinecite{marques2020} features an expanding plasma created by the irradiation of a 4 $\mu$m-thick polypropylene foil with a 500 J, 2 ns-long laser pulse of wavelength $\lambda_0$ = 526 nm and linear polarization. This laser beam is smoothed with a random phase plate and focused by a lens of $f$-number 8, resulting in a focal spot with a speckle pattern of average intensity $\sim 8\cdot 10^{14}$ W/cm$^2$ in the central region, while the maximum speckle intensity is $\sim 5\cdot 10^{15}$ W/cm$^2$. Speckles with typical transverse sizes of a few microns are distributed over a spot with a full width at half maximum of $\sim 150$ $\mu$m. The laser heats the foil up while it is expanding. When the electron density has decreased down to $n_e \sim 0.1 n_c$, the plasma opacity is low enough, and the wave damping is weak enough to generate the Raman instability.

To study numerically these experimental conditions, we employ two-dimensional (2D) PIC simulations \cite{birdsall} (with the code \textsc{Smilei} \cite{derouillat2018}) which describe self-consistently the plasma interaction with electromagnetic fields. The detailed simulation parameters are given in Appendix~\ref{app-simulations}.

\begin{figure}
    \includegraphics[width=8.5cm]{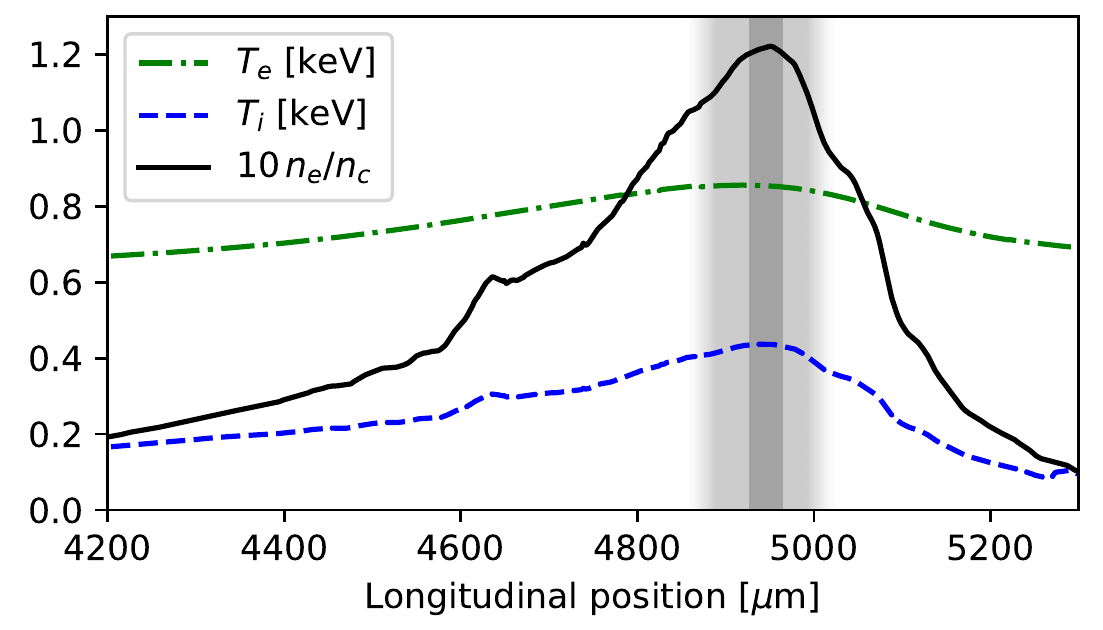}
    \caption{\label{fig-hydro}Density and temperature profiles from hydrodynamic simulations, 1.5 ns after the beginning of the interaction. The light-grey area exemplifies the typical field of view of the experimental diagnostics. The darker grey area corresponds to the portion included in the PIC simulations.}
\end{figure}

For the initial state of the plasma, we rely on the same hydrodynamic simulations as those presented in Ref.~\onlinecite{marques2020}. These used the code FLASH \cite{fryxell2000,tzeferacos2012} in a two-dimensional, cylindrical geometry, with a 7 mm $\times$ 3.5 mm box, a resolution of 0.4 $\mu$m, and 6 levels of mesh refinement. The radiative transfer is solved in a 40-multigroup diffusion approximation. Longitudinal profiles of density and temperature along the central axis were extracted 1.5 ns after the beginning of the interaction, defining the baseline configuration for the PIC simulations. Figure~\ref{fig-hydro} presents these profiles and highlights the portion included in the PIC simulations.

\begin{figure}
    \includegraphics[width=\columnwidth]{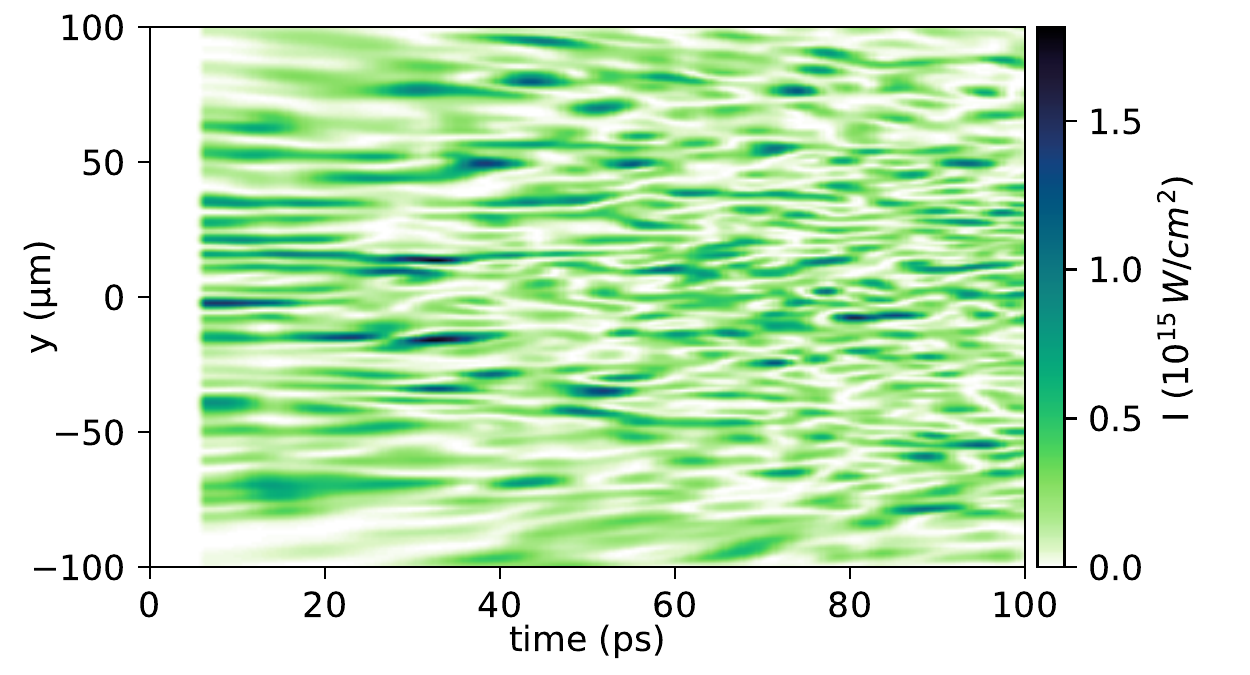}
    \caption{\label{fig-hera}Transverse intensity pattern of the laser at the center of the plasma \textit{vs.} time, as calculated by the code Hera.}
\end{figure}

Being highly resource-intensive, the PIC simulations are limited to a short window, spatially and temporally, assumed to be representative of the whole interaction. These technical limitations (in addition to those from the experiment) prevent a quantitative comparison to the experimental results, but enable qualitative discussions and provide profound insights into the regimes of interaction. Spatially, we reproduce the typical size of a single speckle. Temporally, we have considered simulation durations of the order of the typical speckle life-time. This life-time has been evaluated using the Hera paraxial code \cite{loiseau2006,ballereau2007} which can simulate the whole laser propagation: its intensity speckle pattern (from the random phase plate) is modified by damping (inverse Bremsshtahlung) and by the ponderomotively-driven plasma hydrodynamic evolution, potentially leading to self-focusing and plasma-induced smoothing. The 2D Hera simulation features a 1 mm-long (1000 cells), 0.5 mm-wide (2048 cells) plasma given by 2D hydrodynamic simulations, after plasma creation and heating. The laser focus is located at the center of our simulation box with a $6\cdot 10^{14}$ W/cm$^2$ average intensity. Figure \ref{fig-hera} confirms a non stationary behavior corresponding to plasma-induced smoothing \cite{schmitt1998,grech2006}. The resulting time scale for speckle duration ranges from 5 to 10 ps: this is typically what we use for the PIC simulations.

Concerning the PIC simulations, two laser intensities are presented: either $\sim 6\cdot 10^{14}$ W/cm$^2$ or $\sim4\cdot 10^{15}$ W/cm$^2$, corresponding to a maximum normalized laser vector potential of $a_0=0.016$ or $a_0=0.04$, respectively. Varying the laser intensity corresponds to different scenarios since it is closely related to the amplitude $\delta n$ of the fluctuations of the electron density. This, in turn, influences the growth rate of the LDI \cite{robinson1997}, which is generally proportionnal to $\delta n$.

The fully-ionized, neutral C$_3$H$_8$ plasma has an electron density $n_e\sim 0.122 n_c$ with a profile along the propagation axis $x$ extracted from the same 2D hydrodynamic simulations as discussed above. As the experimental plasma extent is much larger than the PIC simulation box, we limit this profile to the region around the maximum density (down to $\sim 5\%$ below). This is justified by the Raman instability being most effective where gradients are small, and by the experimental diagnostics collecting most of the signal from that central region. The temperatures are taken from the same hydrodynamic simulations which indicate an electron temperature of $T_e=850$~eV and an ion temperature of $T_i=430$~eV.

The dimensionless physical parameters common to all simulations are: $k_\LW \lambda_D \sim 0.17$ (which dictates the damping of LWs), $k_\LWC / k_\LW \sim 0.4$ (which relates to the matching conditions for wave-vectors) and $T_e/T_i \sim 2$ (which is involved in the IAW damping). They are well within the range of conditions obtained from the experimental measurements and from the hydrodynamic simulations.

Some elements are not included in our simulations, such as collisions or ionisation, but we expect those to have very limited impact in the experimental conditions. Note that we verified that including collisions in PIC simulations did not change significantly the results; this is in agreement with the estimated mean free path of one or several hundreds of microns being significantly larger than the size of a speckle. Moreover, a three-dimensional simulation might reveal some differences related to laser self-focusing or polarization effects. This should not affect the general picture.

\section{Simulation at low power\label{sec-loP}}

\begin{figure}
    \includegraphics[width=\columnwidth]{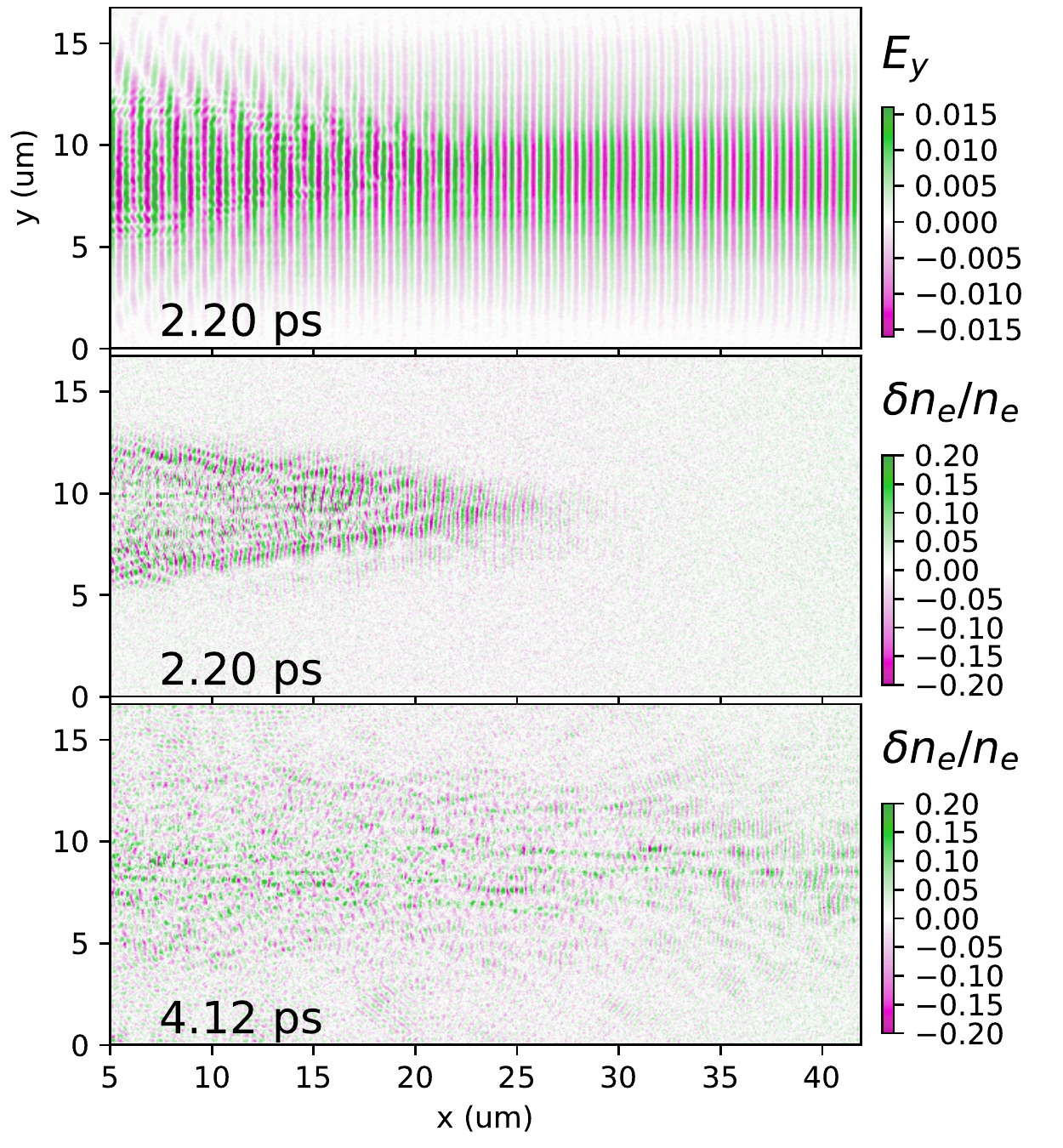}
    \caption{\label{fig-maps1}Low-intensity case. First panel: typical map of the transverse electric field. Other panels: fluctuations of the electron density at two different times representative of the growth and turbulent phases.}
\end{figure}

In a first configuration, we set the laser speckle to a relatively low intensity of $\sim 6\cdot 10^{14}$ W/cm$^2$ ($a_0=0.016$). This intensity is representative of the average focal-spot intensity in the experiment. Figure \ref{fig-maps1} (first panel) illustrates the transverse electric field (mainly the incident laser) that remains fairly uniform after the backscattering has started. The other panels show the density fluctuations associated to all electrostatic waves at two different times: during the SRS linear growth (depending on the location in space) and after the whole box has become turbulent.

\begin{figure}
    \includegraphics[height=11cm]{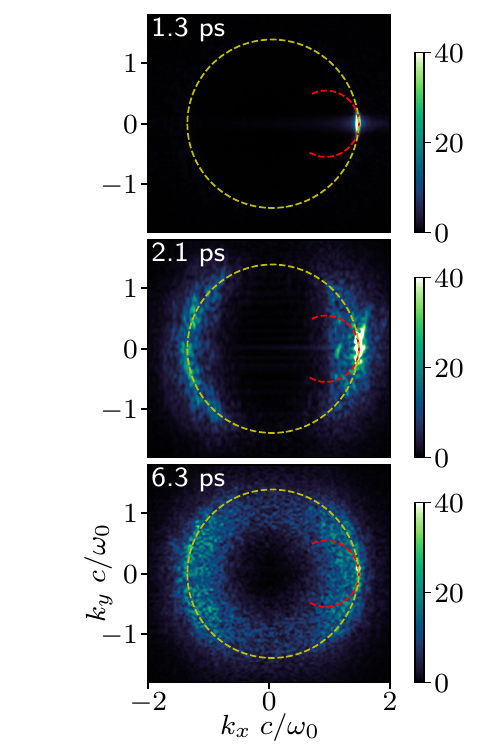}
    \caption{\label{fig-EPW1}Low-intensity case. Spatial spectra of LWs ($d^2\tilde n_e/dk_xdk_y$) in arbitrary units at three different times. Red and yellow lines identify the wave-vectors produced by SRS (Eq. \ref{eq-ramanLW}) and by the first LDI step (Eq. \ref{eq-LDILW}), respectively.}
\end{figure}

We performed 3D (space \& time) Fourier transforms on these electron density maps, over 2D space and a given range of time ($\sim 100\,\omega_0^{-1}\sim 0.03$ ps). By selecting the wave-vectors ($k_x$, $k_y$) and frequencies that satisfy the LW dispersion relation (keeping only positive frequencies), we obtain $k_x$-$k_y$ spatial spectra at different times. Figure \ref{fig-EPW1} shows examples at three different times. Red and yellow lines identify the Raman- and LDI-produced waves, respectively. The Raman peak (at $k_x\sim 1.4\omega_0/c$, consistently with Eq.\ref{eq-ramanLW}) is initially very strong but weakens gradually. LDI-produced waves are initially propagating backwards but the scattering cascade progressively isotropizes their direction. Note that the change in wave-vector length is only of a few percent for each cascade, so that the spectrum does not fill the low-$k$ region.

\begin{figure}
    \includegraphics[height=11cm]{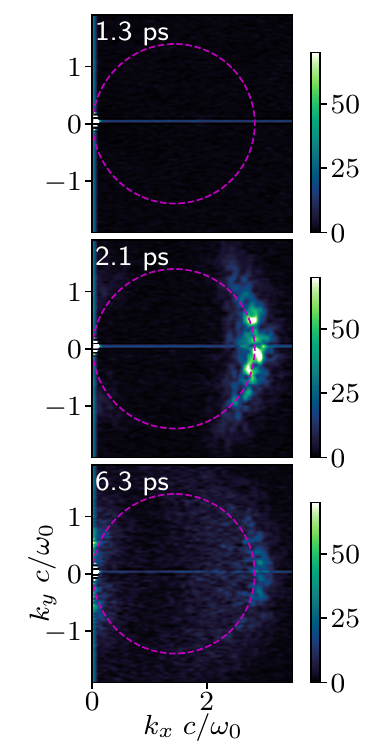}
    \caption{\label{fig-IAW1}Low-intensity case. Spatial spectra of IAWs ($d^2\tilde n_i/dk_xdk_y$) at the same times as Fig. \ref{fig-EPW1}. Magenta lines identify the wave-vectors produced by LDI (Eq. \ref{eq-LDIIAW}).}
\end{figure}

To ensure that these LWs are associated to IAWs with the expected wave-vectors, $x-y$ 2D Fourier transforms have been performed on the ion density maps (given the collection time considered, the low frequency of the ion acoustic modes prevents any distinction between positive and negative frequencies; this makes 3D Fourier transforms superfluous). Examples are given in Fig. \ref{fig-IAW1} at three different times. The IAW wave-vectors match the expected form of Eq. \ref{eq-LDIIAW}. Their angular distribution is less isotropic than that of LWs, because, due to momentum conservation, their angle is approximately twice smaller.

\begin{figure}
    \includegraphics[width=7cm]{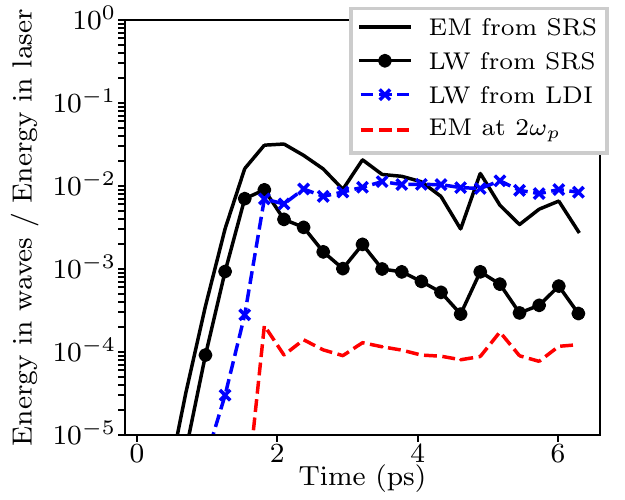}
    \caption{\label{fig-history1}Low-intensity case. Energy in various waves as a function of time, with respect to the current laser energy in the simulation box.}
\end{figure}

For a quantitative comparison, Fig.~\ref{fig-history1} shows the evolution of the total energy in different waves. This is obtained by integrating the $k_x$-$k_y$ spectra of EM waves and LWs over wave-vectors of interest. Note that we distinguish LWs from SRS and those from LDI using the very strong directivity of the former. All waves initially grow exponentially (linear phase) but they saturate after 1.6 ps, due to the large amount of LDI-scattered LWs which deplete the Raman LWs. Whereas the latter were 10 times stronger than the former in the linear phase, they become 10 times weaker 2 ps after saturation. LDI-scattered LWs stay at their saturation level as long as the laser remains on. Overall, they dominate the LW spectrum for most of the simulation duration.

\begin{figure}
    \includegraphics[width=8.5cm]{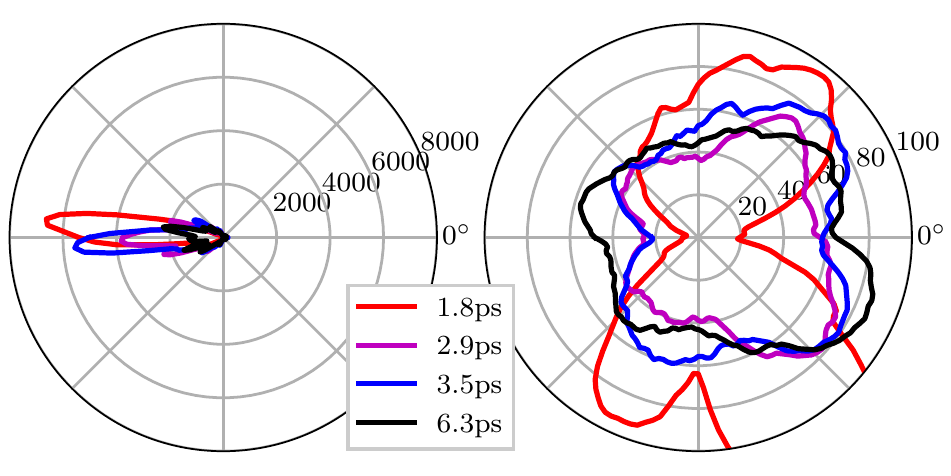}
    \caption{\label{fig-TPC1}Low-intensity case. Angular distribution of EM waves ($|d\tilde B_z/d\theta |^2$). Left: Raman. Right: $2\omega_p$.}
\end{figure}

As a result of SRS-produced LWs being depleted, the LW spectrum loses its initial directivity (close to the laser $\vec k_0$) given by the SRS instability. This depletion also causes LW coalescence to mostly occur between two secondary LDI-scattered waves ($\LW_1$) instead of between one LDI-scattered and one Raman wave ($\LW_1$ and $\LW_0$). These secondary LWs are eventually uniformly distributed in angle, thus the $2\omega_p$ EM radiation will also have no significant directivity. This isotropization of EM radiation will actually occur faster than that of the LWs. Indeed, EM radiation at any angle solely requires $\vec k_{\LW i}$ and $\vec k_{\LW j}$ to differ by $\sqrt{3}\,\omega_p/c$, which corresponds to a $\sim 40\%$ difference: $k$-spectra with 40\% spread (in length or angle) may produce EM waves in all directions. This is confirmed in Fig.~\ref{fig-TPC1} where the angular distribution is given at a few different times. Even though there are preferential emission directions at early times (when SRS-produced waves are still dominant), this is rapidly lost in favor of a more isotropic angular distribution.

Another consequence of the isotropic LW spectrum is that the EM emission will lose its polarization. Indeed, for a given emission direction $\vec e$, the wave-vectors $\vec k_{\LW i}$ and $\vec k_{\LW j}$ of the two parent LWs define a plane containing both $\vec e$ and the polarization axis. As $\vec k_{\LW i}$ and $\vec k_{\LW j}$ become isotropically distributed, the same applies to the polarisation axis, resulting in a less-polarized emission. Note that this effect cannot be observed in the present simulations because they are two-dimensional: LW coalescence can only produce waves polarized in the plane given by their wave-vectors ($x$-$y$ here).

We have performed similar simulations at the same intensity, varying different plasma parameters. The isotropisation is always more pronounced when LDI is more efficient. The most important parameter appears to be the plasma density: lower densities mean less LDI, and, for a density of $n_e=0.07n_c$, the SRS-produced LWs are actually still dominant even after 10~ps. Other parameters that reduce the LDI cascade are: gradients in the density or laser profile, ions with high mass, and high $T_i/T_e$.

\section{Simulation at high power\label{sec-hiP}}

\begin{figure}
    \includegraphics[width=\columnwidth]{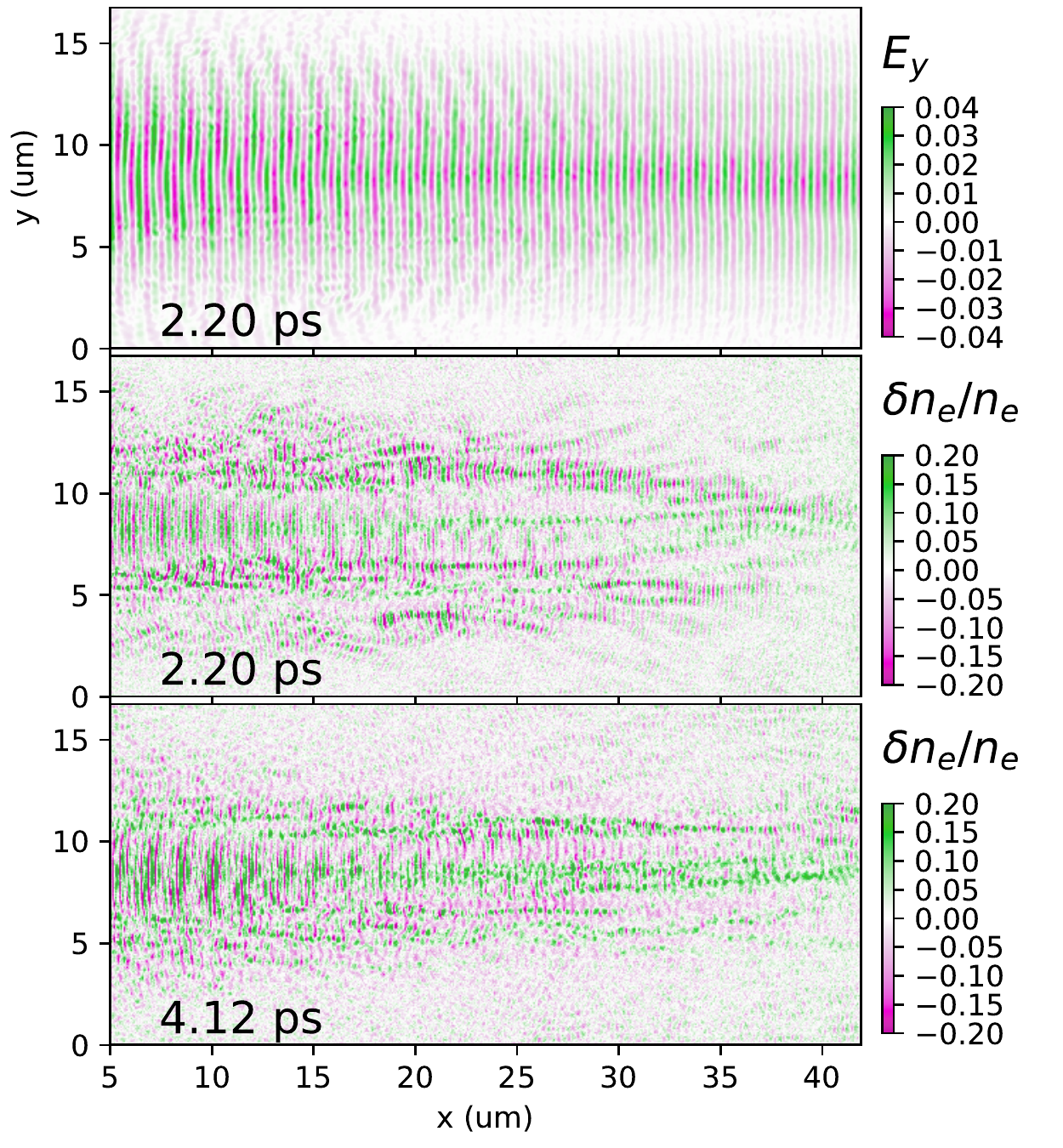}
    \caption{\label{fig-maps2}High-intensity case. First panel: typical map of the transverse electric field. Other panels: fluctuations of the electron density at two different times.}
\end{figure}

We now set the laser speckle to an intensity of $\sim 4\cdot 10^{15}$ W/cm$^2$ ($a_0=0.04$), representative of the strongest speckles in the experiment, where the self-focusing and filamentation of the laser wave are important. This appears in the first panel of Figure \ref{fig-maps2} showing the laser's transverse electric field. It also illustrates the behavior of the electron density: even in the turbulent regime, the strong laser field drives (via SRS) coherent LWs in the form of filaments (strong turbulence).

\begin{figure}
    \includegraphics[height=11cm]{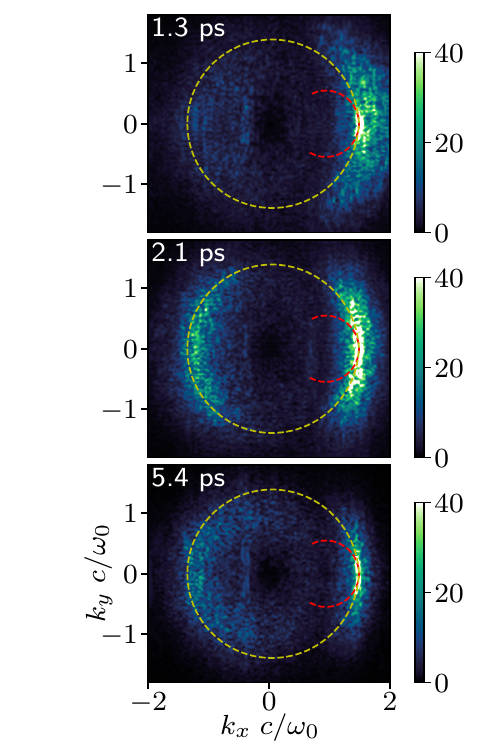}
    \caption{\label{fig-EPW2}High-intensity case. Spatial spectra of LWs ($d^2\tilde n_e/dk_xdk_y$) in arbitrary units at three different times. Red and yellow lines identify the wave-vectors produced by Raman (Eq. \ref{eq-ramanLW}) and LDI (Eq. \ref{eq-LDILW}), respectively.}
\end{figure}

The corresponding LW spectra, shown in Fig. \ref{fig-EPW2}, are obtained in the same manner as in Sec. \ref{sec-loP}. While the LDI-induced LWs grow with negative $k_x$, the Raman waves remain strong. Contrary to the previous low-intensity case, two groups of LWs (forward waves close to the SRS wave-vector and backward waves from LDI) persist at all times. No obvious LDI cascade contributes to their isotropisation, as visible in the last panel of Fig. \ref{fig-EPW2}.

\begin{figure}
    \includegraphics[width=7cm]{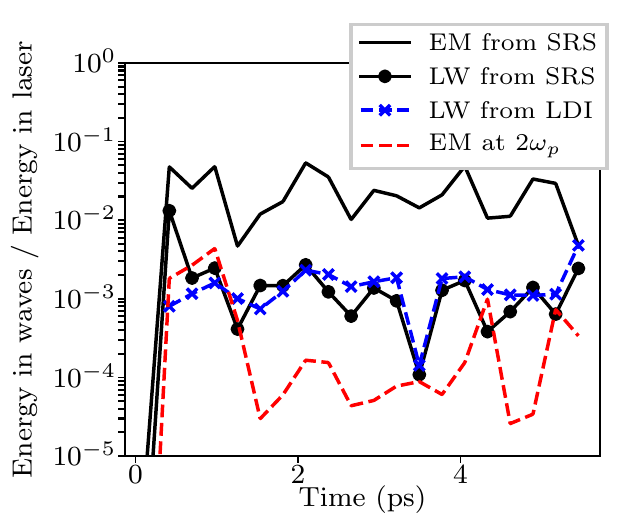}
    \caption{\label{fig-history2}High-intensity case. Energy in various waves as a function of time, with respect to the current laser energy in the simulation box.}
\end{figure}

Figure \ref{fig-history2} provides a quantitative comparison of the energy in the various waves. The Raman EM waves are produced more efficiently than in the low-intensity case, but the more striking difference is that the Raman LWs are not depleted by LDI. They remain constant, at the same level as LWs from LDI, whereas they ended lower by more than one order of magnitude in the previous section.

\begin{figure}
    \includegraphics[width=8.5cm]{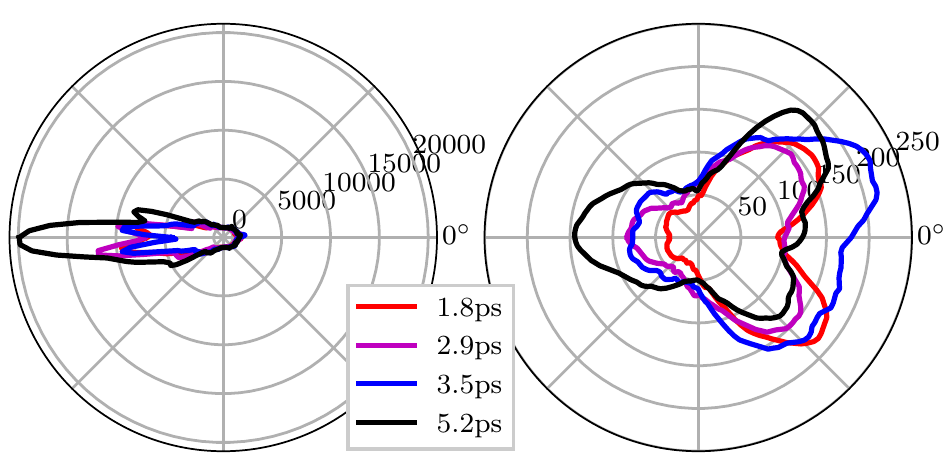}
    \caption{\label{fig-TPC2}High-intensity case. Angular distribution of EM waves ($|d\tilde B_z/d\theta |^2$). Left: Raman. Right: $2\omega_p$.}
\end{figure}

The consequence of LWs being now split in two halves (peaked Raman forward, broad LDI backward) is that the Raman directivity will be imprinted on the angular distribution of the $2\omega_p$ EM waves resulting from LW coalescence. This is illustrated in Fig.~\ref{fig-TPC2}. The direction of these EM waves is mostly forward, by construction of wave-vectors that must satisfy the conservation of momentum. Indeed, the LWs from LDI have a slightly shorter wave-vector than those from SRS, and this constrains their direction. The $2\omega_p$ emission in the opposite direction ($180^\circ$) is explained by another construction of wave-vectors: there is a weaker forward-SRS instability that creates backward LWs with shorter wave-vectors. These LWs are visible in Fig. \ref{fig-EPW2} at $k_x\sim -0.5\, \omega_0/c$, and after isotropization through LDI (not highlighted in the figure) they couple with the initial LWs to produce backwards radiation.

In addition to an angular distribution with some directivity, this high-intensity case would likely exhibit a higher degree of polarization. Indeed, one of the two LWs generating $2\omega_p$ emission has to come directly from SRS, with a narrow forward direction $\vec k_{\LW 0}$. Knowing the emission direction $\vec e$, the polarization plane contains both $\vec k_{\LW 0}$ and $\vec e$, which explains why the degree of polarization will be proportionnal to the directivity of $\vec k_{\LW 0}$.

\section{Comparison to experimental results\label{sec-comparison}}

We have performed a large number of simulations, but have limited this discussion on a comparison between low and high-intensity speckles, while keeping all other parameters as close as possible to the experimental situation of Ref. \onlinecite{marques2020}. The goal was to show two limits in which the LW spectrum becomes either isotropic or directional, and the $2\omega_p$ emission has a corresponding angular distribution and degree of polarization. Note that the experimental results did not provide a direct measurement of the angular distribution: the hydrodynamic simulations proved that strong refraction from the plasma prevented a proper analysis. Indeed, the location of the emitting plasma is not known with sufficient precision to calculate the effect of refraction with sufficient certainty. We will thus discard the measurements at various angles from our analysis, and focus instead our discussion on the conversion efficiency and the polarization.

A first comparison to that experimental study is the estimated energy conversion efficiency from laser to $2\omega_p$ emission, $\sim 1.7\cdot 10^{-6}$ according to the experimental results. To compare with the simulations, we must take into account three factors. (i) The experimental value corresponds to the total, integrated emissions, while they are instantaneous in the simulations. As the laser duration is 2 ns and the emission lasted only 0.3 ns, we must multiply the experimental value by $\sim 7$. (ii) A significant fraction of the laser energy is used in forming and heating the plasma. This fraction $\sim$ 75\%, was directly extracted from the hydrodynamic simulations. (iii) The experimental plasma was large enough to cause some reabsorption. Using the hydrodynamic simulation results, we performed a ray tracing computation to estimate this effect showing that a quarter of the initial emission is absorbed in the surrounding plasma. These three corrections bring the instantaneous, local value for conversion efficiency to $6\cdot 10^{-5}$, well within the simulation results which show an efficiency between $5\cdot10^{-5}$ and $10^{-4}$. Note that the whole laser pattern may not contribute equally to the emission as its spatial distribution, and that of the plasma, are not homogeneous. This factor cannot be precisely calculated as it would require PIC simulations to cover a much larger space, but it would bring the compared efficiency closer.

An intermediate efficiency comparison may also be discussed: the conversion from laser to SRS EM waves at 180$^\circ$. The experimental data shows a typical peak SRS emission of $\sim 10$ GW$/$sr. Its angular spread is not known directly, but it is estimated using two detections at two slightly different angles to be $\sim 0.1$ sr, thus a backward-SRS peak power of $\sim 1$ GW. The laser power is of the order of $200$ GW, which results in a conversion efficiency of a $\sim 5\cdot 10^{-3}$. This is reasonably close to the simulated conversion ($10^{-2}$ up to a few $10^{-2}$).

The polarization of the $2\omega_p$ emission provides a stronger characterization of the experimental scenario. It is related to the LW angular distribution. Indeed, the polarization always lies in the plane formed by the two wave-vectors $\vec k_{\LW i}$ and $\vec k_{\LW j}$, which also contains the direction of observation. As a consequence, the LW angular distribution around the axis of observation determines the degree of polarization. In the situation studied in Sec.~\ref{sec-loP}, the LWs are relatively uniformly distributed (cf. Fig.~\ref{fig-EPW1}). Numerical estimates lead to a reduction of the signal of less than 50\% if a polarizer was placed with its axis along $y$ (compared to a fully non-polarized emission). In the other case (Sec.~\ref{sec-hiP}, Fig.~\ref{fig-EPW2}), we find that the signal would be reduced by an order of magnitude, as the forward LWs dominate the spectrum and favor a polarization along $x$. This second case matches the experimental results where the presence of a polarizer reduced the signal by an order of magnitude, although a quantitative comparison is not possible due to shot-to-shot variations. Nevertheless, the $2\omega_p$ polarization shows that the experimental conditions are overall similar to those presented in the second, high-power case. We can thus conclude that the high-power speckles present in the laser profile contribute to most of the signal.

\section{Conclusions\label{sec-conclusion}}

In summary, PIC numerical simulations have proven succesful in describing the physical scenario for the growth and coupling of waves that ultimately generate $2\omega_p$ emission in a laser-plasma experiment. The rich structure of this setup can have many outcomes, but we identified two main routes which result in different angular distributions and degrees of polarization. They are mainly governed by the growth of the LDI which controls the degree of anisotropy of LWs. The LDI growth, in turn, depends on various parameters (such as the plasma density, the ion and electron temperatures, etc.); we presented two cases differing by the laser intensity as it appeared a predominant parameter which can be easily linked to experimental conditions. With very low wave amplitudes and little LDI growth, turbulence would be very weak, and LWs would remain mostly on-axis, thus yielding directive and polarized emission. As the amplitudes increase, turbulence tends to isotropize IAWs and LWs, causing loss of directivity and polarization, but those are recovered with higher amplitudes that force the system in the input's direction.

The main perspective for this study is to relate these findings to the space-physics conditions, and eventually compare to results from spacecraft or ground observation. Our study shows that the breadth of LW $k$-spectra strongly affects the directionnality and the polarization of the $2\omega_p$ emission. This fact will likely apply to space-plasma configurations. Future experiments will be focused on the differences between the laser and space setups: ratios between wave-vectors, temperatures, and waves amplitudes. A constant magnetic field could also be applied on the plasma to match the interplanetary conditions more closely.

\section*{Supplementary material}

An input file for the Smilei code is provided as supplementary material.

\section*{Acknowledgements}

Simulations were performed using the open-source code Smilei \footnote{{https://github.com/SmileiPIC/Smilei}} on the Joliot-Curie machine hosted at TGCC-France \footnote{{http://www-hpc.cea.fr/en/complexe/tgcc-JoliotCurie.htm}}, using High Performance Computing resources from GENCI-TGCC (Grant No. 2018-x2016057678).

\section*{Data availability}

Raw simulation data were generated at the Joliot-Curie supercomputing facility. Derived data supporting the findings of this study are available from the corresponding author upon reasonable request.

\appendix
\section{Notations\label{app-notations}}

\begin{itemize}

    \item Speed of light in vacuum: $c$
    \item Permittivity of vacuum: $\varepsilon_0$
    \item Permeability of vacuum: $\mu_0$
    \item Electron and ion mass: $m_e$ and $m_i$
    \item Electron charge: $-e$
    \item Ion average charge: $Z\, e$
    \item Laser frequency: $\omega_0$
    \item Laser wave-vector in the plasma: $k_0$
    \item Laser critical density: $n_c = \varepsilon_0 m_e \omega_0^2 / e^2$
    \item Laser normalized amplitude: $a_0 = \sqrt{\frac{\mu_0 e^2 I \lambda_0^2}{2\pi^2 m_e^2 c^3}}$, where $I$ is the average laser intensity
    \item Electron and ion temperatures (in units of energy): $T_e$ and $T_i$
    \item Electron and ion densities: $n_e$ and $n_i$
    \item Thermal velocity: $v_\textrm{th}=\sqrt{T_e/m_e}$
    \item Ion acoustic velocity: $c_s\sim\sqrt{(Z T_e + 3 T_i)/m_i}$
    \item Electron plasma frequency: $\omega_p=\sqrt{\frac{n_e e^2}{m_e \varepsilon_0}}$
    \item Electron Debye length: $\lambda_D = v_\textrm{th} / \omega_p$
    \item $\eta = \omega_p/\omega_0 = \sqrt{n_e/n_c}$
    
\end{itemize}

\section{Simulation parameters\label{app-simulations}}

Particle-in-cell simulations describe the plasma particle distribution as a collection of discrete \textit{macro}-particles. The electromagnetic fields are computed on regular grids. Fields affect macro-particless through the Lorentz force, and macro-particles affect fields via their current in Maxwell's equations. In this article, all PIC simulations are carried out using the open-source code \textsc{Smilei} \cite{derouillat2018}. Maxwell's equations are solved with the finite-difference time-domain method using a staggered Yee grid \cite{nuter2014}, and the projection of particle currents on the grid follows Esirkepov's algorithm\cite{esirkepov2001}.

The PIC simulations are 2D in space (3D in momentum space) with a cartesian domain of size $80\, \lambda_0$ by $32\, \lambda_0$ with $3072\times 1280$ cells, and a timestep of $0.1\,\omega_0^{-1}$. Spatially, we mimic the transverse shape of a single speckle: the laser transverse envelope has a gaussian shape with a full-width at half-maximum of $12\lambda_0$. The laser field's temporal envelope is a short (100 fs) linear ramp up to a constant value. Importantly, the boundary conditions are absorbing in all directions so that the side-scattered $2\omega_p$ electromagnetic waves do not cycle back into the box and cause spurious signal.

Each particle species (C, H, and electron) of the plasma has 196 macro-particles per cell. The total number of particles in one simulation is $\sim2\cdot 10^9$. They are split between 64 nodes (1536 cores) of the Irene-Skylake partition of the Joliot-Curie supercomputer. The typical time for one core to process one timestep for one particle is 180~ns. Overall, one simulation uses 25000 to 50000 core-hours for a physical time of $\sim 10$ ps.

%
  
\end{document}